\documentclass[review]{elsarticle}

\usepackage{lineno,hyperref}
\usepackage{caption}
\usepackage{subcaption}
\usepackage{comment}
\usepackage{makecell}
\usepackage{tikz}

\journal{arXiv}

\makeatletter
\def\@author#1{\g@addto@macro\elsauthors{\normalsize%
    \def\baselinestretch{1}%
    \upshape\authorsep#1\unskip\textsuperscript{%
      \ifx\@fnmark\@empty\else\unskip\sep\@fnmark\let\sep=,\fi
      \ifx\@corref\@empty\else\unskip\sep\@corref\let\sep=,\fi
      }%
    \def\authorsep{\unskip,\space}%
    \global\let\@fnmark\@empty
    \global\let\@corref\@empty  
    \global\let\sep\@empty}%
    \@eadauthor={#1}
}
\makeatother

\def\checkmark{\tikz\fill[scale=0.4](0,.35) -- (.25,0) -- (1,.7) -- (.25,.15) -- cycle;}

\begin{document}
\nolinenumbers

\begin{frontmatter}

\title{Potential energy curves of molecular nitrogen up to N$_2^{4+}$}

\author{A. Hadjipittas \corref{cor1}\fnref{fn1}}
\ead{a.hadjipittas.16@ucl.ac.uk}
\cortext[cor1]{Corresponding author}

\author{A. Emmanouilidou\fnref{fn1}}

\address{$^1$ Department of Physics and Astronomy, University College London, Gower Street, London WC1E 6BT,  England, United Kingdom}



\begin{abstract}
The potential energy curves for molecular ions up to N$_2^{4+}$ are calculated in an ab initio manner using the multi configurational self-consistent field method.  Specifically, we implement in an automatic way a previously used double loop optimisation scheme within the multi configurational self-consisted field method.  We obtain the potential energy curves up to N$_2^{4+}$ ions with any combination of core, inner valence, and outer valence holes. Finally, we provide the code used to generate these potential energy curves. 
\end{abstract}

\end{frontmatter}

\section{Introduction}

Obtaining potential energy curves (PECs) for molecular ions provides one way to account for nuclear dynamics, in addition to electronic processes, when molecules interact with VUV and XUV laser pulses \cite{Deb6,Deb7}.  The theoretical description of the interplay of inner shell electronic processes with nuclear motion is timely due to the advancements in generating VUV and XUV pulses with free-electron lasers (FEL) \cite{Deb2, Deb3}. FEL pulses interacting with molecules result in ionisation of inner-shell electrons via a single-photon absorption or a multi-photon one.  A molecule with an inner shell hole will relax to a more stable state via one or more Auger processes.  An Auger process occurs when an electron falls from a higher-energy orbital filling in an existing lower-orbital hole. The energy released from this transition is enough to ionise one or more bound electrons. The interplay of photo-ionisation and Auger processes gives rise to a plethora of possible pathways that lead to the formation of a certain molecular ion \cite{Wallis2014, Young2010}.  To accurately describe the interplay between photo-ionisation and Auger processes in molecules, it is necessary to also account for nuclear motion.  In previous studies, nuclear motion was included mostly in a phenomenological way via rate equations \cite{Deb11,Banks2017,Liu2016}, while there are few studies where nuclear motion is included \cite{Arnold_Santra_2017,Li_Santra_2015}. To account for nuclear motion beyond a phenomenological description, we need to obtain accurate PECs for molecular ions with any kind of holes, including inner shell holes. \newline

A number of theoretical studies have previously computed the PECs of singly and doubly ionised states of N$_2$ with one or two valence electrons or one core electron missing \cite{Nagy_1999,Polak_2003,Gagnon_2007,Aoto_2006,Trabbatoni_2015} .  The computations in these studies were done using the multiconfigurational self-consistent field method (MCSCF) \cite{Roothaan_1979,Werner_1981,Lengsfield_1981}. Other theoretical studies, using different techniques, have computed the PECs of N$_2$ states with an inner valence hole \cite{Deb24} or a single core hole \cite{Deb25,Deb26,Deb27}. The difficulty in calculating PECs of N$_2$ ions with inner valence and/or core holes is that there are other states with the same symmetry but lower energy. This leads to variational collapse \cite{Besley_2009} to the lowest energy state during optimisation of the molecular orbitals and the respected coefficients. Previous studies \cite{Deb25,Deb26,Deb27} employed a two-step optimisation process in order to avoid variational collapse when computing the PECs of single core hole states. This two-step optimisation process was introduced by Rocha et al. \cite{Deb30} and it entails freezing the core orbitals in the first step while optimising the valence orbitals using the MCSCF method. In the second step, one freezes the valence orbitals and optimises the core orbitals. A similar two-step optimisation process was used by Carravetta et al. \cite{Carravetta_2013} to compute the energy of a number of N$_2$ ions with multiple core holes at equilibrium nuclear distances.  More recently, in the work of Bhattacharya et al. \cite{Deb}, the two-step optimisation process was used to compute the PECs for a number of N$_2$ ions with up to two electrons missing as a function of nuclear distance.  \newline

In addition, in reference \cite{Deb} three techniques were used in order to compute the PECs of N$_2$ ions with any combinations of up to two electrons missing.  These calculations were carried out using the CASSCF method \cite{Kreplin_2019,Kreplin_2020} within the framework of the quantum chemistry package Molpro \cite{Werner_2012,Werner_2020}.  The CASSCF method was performed using an active space of 10 orbitals, i.e. the seven doubly occupied orbitals of the ground state of N$_2$, $1\sigma_{g}^2 1\sigma_{u}^22\sigma_g^2 2\sigma_u^2 1\pi_{u_{x}}^2 1\pi_{u_{y}}^2 3\sigma_{g}^2$ and three more virtual orbitals, which are the lowest energy virtual ones. In the first technique, a single optimisation is carried out using CASSCF in order to compute the PECs of states with one or two valence holes that have the lowest energy of their respective symmetry.  In the second technique, the state-averaging CASSCF (SA-CASSCF) is employed in Molpro to compute PECs of singly ionised N$_2$ states with one inner valence hole, or doubly ionised N$_2$ states with two outer valence holes. This second technique is used to obtain the above mentioned singly or doubly ionised N$_2$ states that,  however, do not have the lowest energy of their respective symmetry. In this state-averaging technique, the wavefunctions of the resulting states are expressed as a linear combination of the wavefunctions of different electronic configurations. We map each of these averaged states to the electronic configuration that has the biggest overlap with the averaged state at the equilibrium internuclear distance. In the third technique, the two-step optimisation process \cite{Deb30} is employed to compute the PECs of states with one or two core electrons, or one inner valence and one more any other kind of electron, missing. The authors in reference \cite{Deb}, freeze the core and/or inner valence orbitals corresponding to the electrons missing and optimise the remaining orbitals in the first step. In the second step, the previously frozen orbitals are now optimised. \newline

In order to generate PECs of all ions  up to N$_2^{4+}$, we generalise the three techniques used by Bhattacharya et al. \cite{Deb} in the context of Molpro. In addition to the three techniques previously employed in Ref. \cite{Deb}, we use a fourth technique to compute the PECs of certain states of N$_2^{3+}$ and N$_2^{4+}$. These states have at least one core electron missing and do not have the lowest energy of their respective symmetry. This fourth technique combines the two-step optimisation process with the state-averaging technique. \newline 

In this work, we also provide a python code \cite{github} that automatically generates the input Molpro files that are necessary to compute the PECs for any combination of up to N$_2^{4+}$ ion states with any combination of orbital vacancies. In addition, the python code we provide outputs the wavefunctions of all orbitals for any ion state up to N$_2^{4+}$ for any internuclear distance. This additional feature can be used to obtain single-photon absorption cross sections and Auger decay rates when the nuclear and electronic motion are both accounted for during the interaction of N$_2$ with a VUV or XUV laser pulse. The techniques and python code provided in this work can be adjusted to obtain the PECs for other diatomic molecules. 

\section{Method}
\label{sec:Method}
In the following, we describe the computation of the PECs of up to N$_2^{4+}$ ion states. The multi-configurational self-consistent field method (MCSCF) combines the self-consistent field method with the configuration interaction method \cite{Deb33}. MCSCF uses a combination of Slater determinants that account for all electronic excitations of the molecular ion under consideration. In this work, we employ the complete active space variant of MCSCF, known as CASSCF \cite{Deb37,Deb38}. We consider a total of 10 active orbitals in our computations. These are the seven doubly occupied orbitals of the N$_2$ ground state $1\sigma_{g}^2 1\sigma_{u}^22\sigma_g^2 2\sigma_u^2 1\pi_{u_{x}}^2 1\pi_{u_{y}}^2 3\sigma_{g}^2$ and the three virtual orbitals $3\sigma_u$,  $1\pi_{g_{x}}$ and $1\pi_{g_{y}}$. In this work, the orbitals $1\sigma_{g}$ and $ 1\sigma_{u}$ are referred to as core orbitals,  orbital $2\sigma_g$ is referred to as inner valence orbital and the rest of the doubly occupied orbitals are referred to as outer valence orbitals. We employ the augmented Dunning correlation consistent quadruple valence basis set (aug-cc-pVQZ) \cite{Deb43} for all our calculations. We carry out our calculations using the standard release version MOLPRO2020.1 of Molpro \cite{Werner_2012,Werner_2020}. We initially describe the orbitals of the N$_2$ ion under consideration using the Hartree-Fock orbitals of the ground state of N$_2$. We improve this initial description of the orbitals of the N$_2$ ion under consideration by performing a CASSCF calculation of the ground state of N$_2$. These CASSCF orbitals are the input for all methods that we use in what follows. \newline

We use four techniques to calculate the PECs of interest. The first one involves a single optimisation of all ten orbitals using CASSCF. The second one employs the state-averaging CASSCF (SA-CASSCF), while the third technique employs the two-step optimisation process using CASSCF (TS-CASSCF) \cite{Deb25, Deb26, Deb27, Deb30, Carravetta_2013, Deb}. The fourth technique combines state averaging with the two-step optimisation process using CASSCF and we refer to it as SA-TS-CASSCF. In table \ref{tab:techniques}, we show which of the four techniques is employed to compute the PECs of a given N$_2$ ion state up to N$_2^{4+}$.

\begin{center}
\begin{table}[h!]
\centering
\begin{tabular}{|c|c|c|c|c|c|}
\hline
& & \multicolumn{4}{|c|}{Technique used}  \\
 \hline
 & Missing Electrons &  CASSCF & SA-CASSCF & TS-CASSCF & SA-TS-CASSCF \\ 
 \hline
N$_2^{+}$ & 1 Outer   & \checkmark  & & & \\
 \hline
& 1 Inner &  & \checkmark  & &\\
 \hline
& 1 Core &  &  & \checkmark & \\
 \hline
N$_2^{2+}$ & 2 Outer & \checkmark & \checkmark & & \\
 \hline
& 2 Inner & &  & \checkmark & \\
 \hline
& 1  Inner \& 1 Outer &  & & \checkmark & \\
 \hline
& 2 Core & &  & \checkmark & \\
 \hline
& 1 Core \& 1 Outer &  & & \checkmark & \\
 \hline
& 1 Core \& 1 Inner &   && \checkmark & \\ 
 \hline
N$_2^{3+}$ & 3 Outer & \checkmark & \checkmark &  & \\
\hline
& 2 Inner \& 1 Outer & &  & \checkmark & \\
\hline
& 1 Inner \& 2 Outer &  &  &  \checkmark & \\
\hline
& 3 Core &  &  & \checkmark & \\  
\hline
& 2 Core \& 1 Outer/Inner &  &  & \checkmark & \\  
\hline
& 1 Core \& 2 Outer/Inner  &  &  & \checkmark & \checkmark \\  
 \hline 
N$_2^{4+}$ & 4 Outer &  \checkmark & \checkmark & & \\
 \hline
& 2 Inner \& 2 Outer & & & \checkmark & \checkmark \\
\hline
& 1 Inner \& 3 Outer & &  & \checkmark & \checkmark \\
\hline
& 4 Core & & & \checkmark & \\  
\hline
& 3 Core \& 1 Outer/Inner & &  & \checkmark & \\ 
\hline
& 2 Core \& 2 Outer/Inner &  & & \checkmark & \checkmark \\   
\hline
& 1 Core \& 3 Outer/Inner &  & & \checkmark & \checkmark \\   
\hline
\end{tabular}
\caption{We list the optimisation technique employed to obtain the PECs for every possible N$_2$ ion up to N$_2^{4+}$. We denote the outer valence electrons as Outer, the inner valence electrons as Inner and the core electrons as Core. The number denotes the number of missing electrons.}
\label{tab:techniques}
\end{table}
\end{center}

To obtain the PECs for the ion states with at least one core electron missing, or with an inner valence and any other electron missing we use the TS-CASSCF technique or the SA-TS-CASSCF technique. The former technique is used when the state of interest has the lowest energy of its respective symmetry and the latter one otherwise.  In the two-step optimisation, if a core electron is missing we freeze both core orbitals in the first step and optimise them in the second step. We also restrict the electron occupation number to be equal to the actual number of electrons occupying the orbital. For ion states with two or more electrons missing with one of them being an inner valence electron, we freeze the inner valence orbital in the first step and optimise it in the second step. If the number of electrons in the inner shell orbital is one we allow the occupation to vary between zero and one, while if it is zero we restrict the occupation to be equal to zero. During the two-step optimisation process we generally do not freeze the virtual orbitals. However, if one of the orbitals we freeze in the first step has the same symmetry as one of the virtual orbitals, in the second step we also freeze this specific virtual orbital to avoid orbital rotations. In addition, to obtain the PECs for ion states with more than one outer valence electrons missing, we use either the CASSCF or the SA-CASSCF technique depending on whether the ion state has or does not have the lowest energy of its respective symmetry.

\section{Results and Discussion}
\label{sec:Results}
In what follows follows, we obtain the PECs of N$_2$ ion states with up to four electrons missing. First, we compare our results for the PECs of the singly and doubly ionised states of N$_2$ with the PECs obtained in Ref. \cite{Deb}. Then, we compute the PECs of the triply and quadruply ionised states of N$_2$. Our calculations are performed for nuclear distances ranging from 0.5 to 4 \AA. 

\subsection{PECs for singly and doubly ionised states of N$_2$}
\label{sec:PECsComparison}
\begin{figure}[h!]
     \centering
     \includegraphics[width=\textwidth]{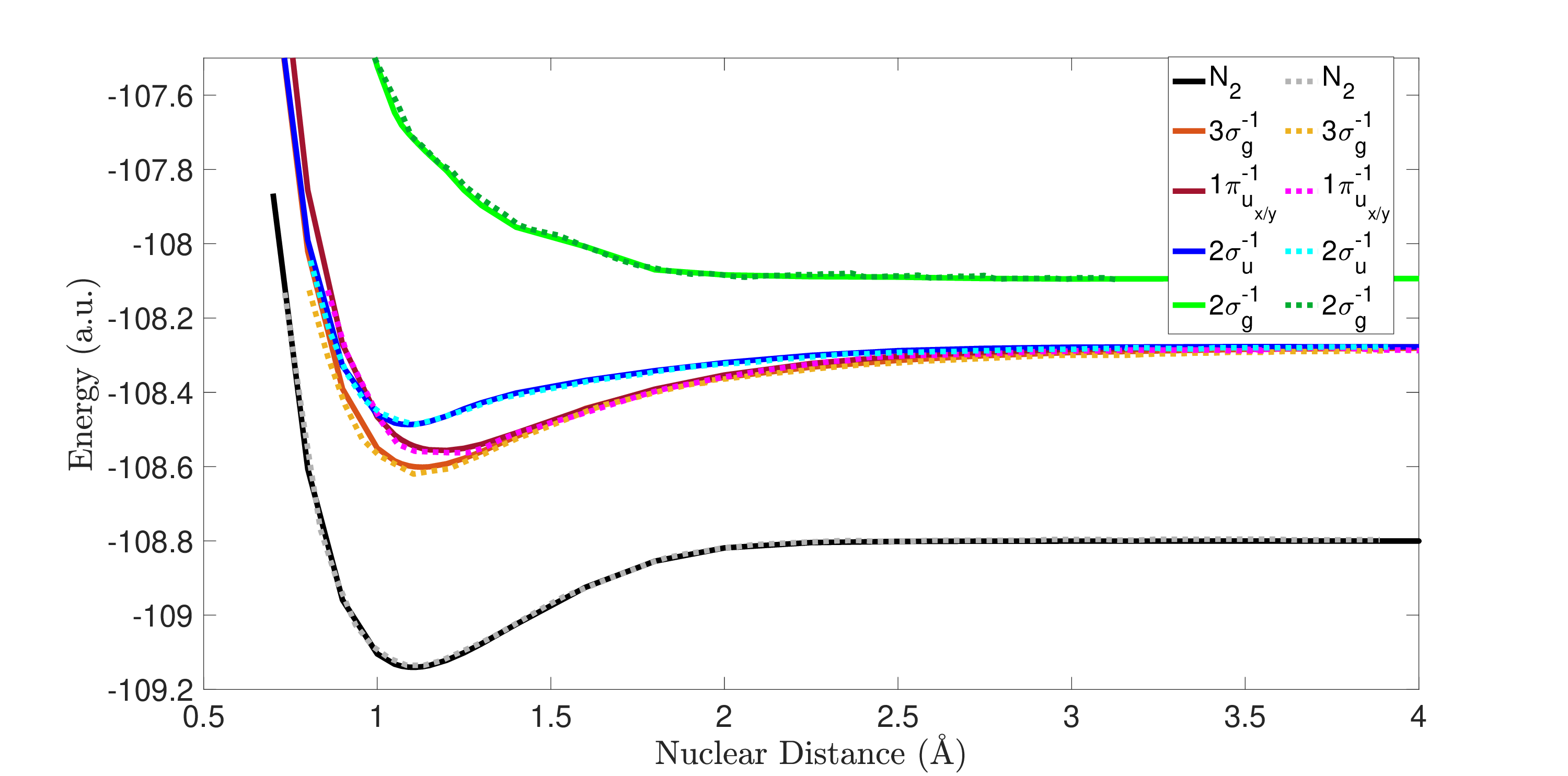}
     \caption{PECs of N$_2^+$ ion states with an inner or outer valence electron missing. Solid curves represent the results obtained in this work, while dotted curves represent the results obtained in Ref. \cite{Deb}. The 1$\pi_{u_{x/y}}^{-1}$ stands for an electron missing from either the 1$\pi_{u_x}$ or the 1$\pi_{u_y}$ orbital.}
     \label{fig:Comp_1+_Outer_Inner}
\end{figure}

In Fig. \ref{fig:Comp_1+_Outer_Inner}, we present the PECs for N$_2^+$ ions with an inner or an outer valence electron missing. The solid black curve corresponds to the PEC of neutral N$_2$. Also, we obtain the PECs of the ion states that are obtained by removing an electron from the 3$\sigma_g$ (solid orange), 1$\pi_{u_{x/y}}$ (solid pink), 2$\sigma_u$ (solid blue) and 2$\sigma_g$ orbital (solid green). We find our results to be in very good agreement with the ones obtained in Ref. \cite{Deb}. The PECs of the ion states having an outer valence electron missing (3$\sigma_g^{-1}$, 1$\pi_{u_{x/y}}^{-1}$ and 2$\sigma_u^{-1}$) were calculated using the CASSCF technique, whereas the state involving an inner valence electron missing (2$\sigma_g^{-1}$) was calculated using the SA-CASSCF technique, see Table \ref{tab:techniques}.  \newline

\begin{figure}[h!]
     \centering
     \includegraphics[width=\textwidth]{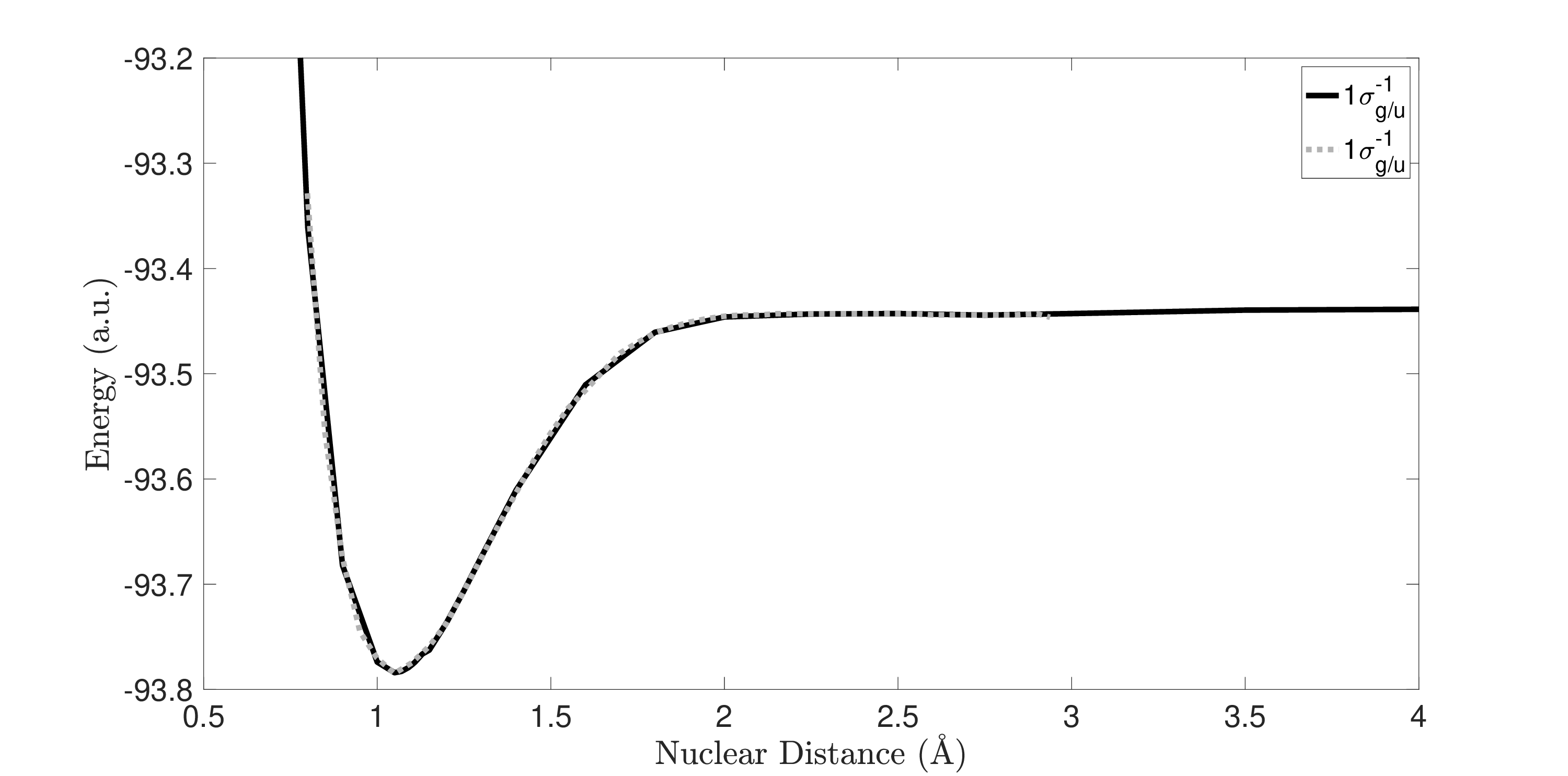}
     \caption{PECs of N$_2^+$ ion states with a core electron missing. The solid curve represents the results obtained in this work, while the dotted curve represents the results obtained in Ref. \cite{Deb}. The PEC of the 1$\sigma_{g/u}^{-1}$ state, accounts for the two distinct PECs of the 1$\sigma_g^{-1}$ and 1$\sigma_u^{-1}$ states, which we find to be very similar.}
     \label{fig:Comp_1+_Core}
\end{figure}

Next, we compute the PECs of N$_2$ ion states with a single core electron missing. For these states, we find that our results for the PECs are in very good agreement with the ones obtained in Ref. \cite{Deb}, see Fig. \ref{fig:Comp_1+_Core}. Our results are denoted with solid curves while the results of Ref. \cite{Deb} are indicated with dotted curves. We find that the PECs corresponding to the ion states with a 1$\sigma_g$ or 1$\sigma_u$ electron missing are almost identical, see Fig. \ref{fig:Comp_1+_Core}. We use the TS-CASSCF technique to obtain the PECs for these two states, see Table \ref{tab:techniques}.  \newline

\begin{figure}[h!]
     \centering
     \includegraphics[width=\textwidth]{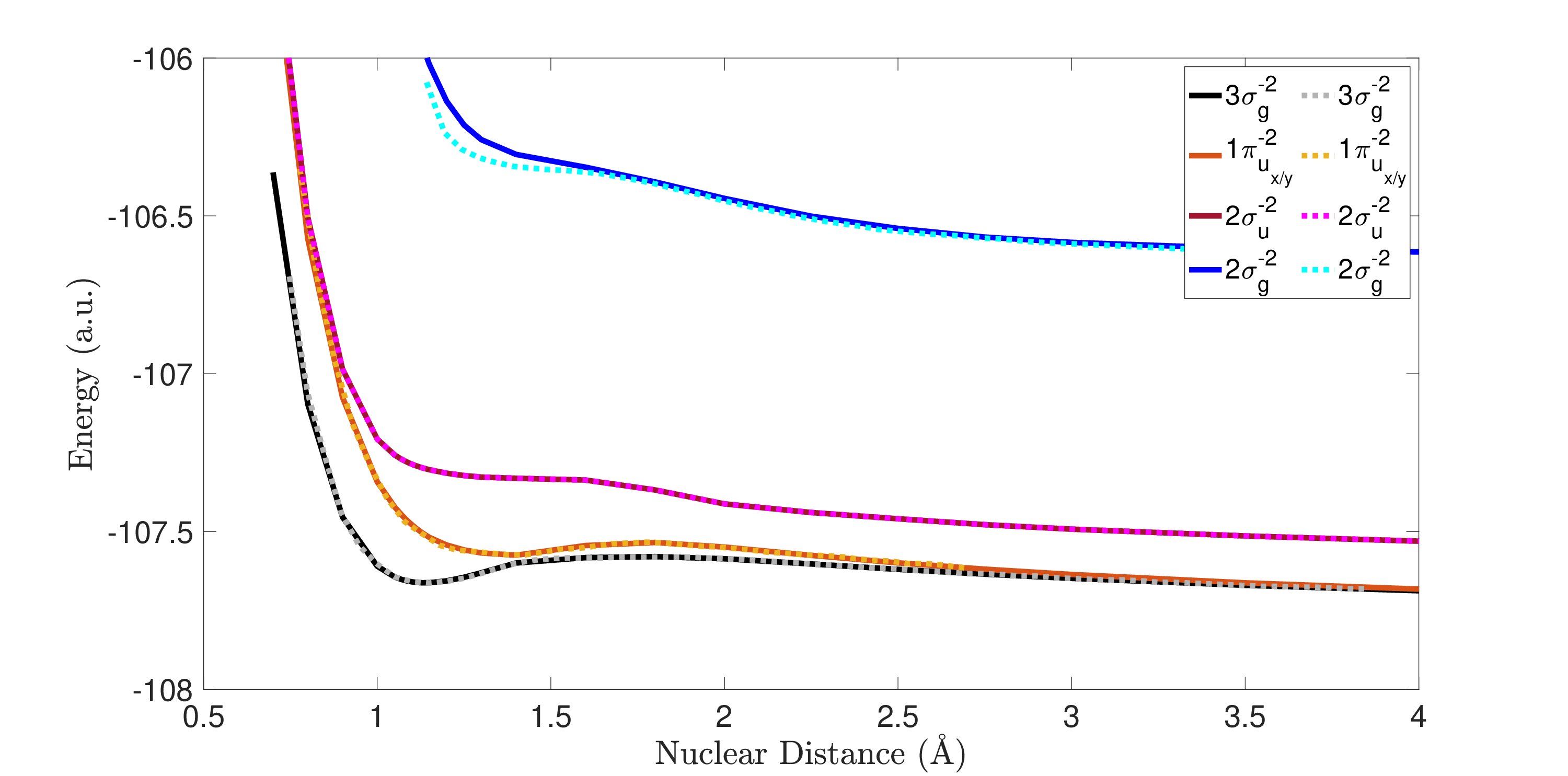}
     \caption{PECs of N$_2^{2+}$ ion states with two outer or two inner electrons missing from the same orbital. Solid curves represent the results obtained in this work, while dotted curves represent the results obtained in Ref. \cite{Deb}. The 1$\pi_{u_{x/y}}^{-2}$ stands for two electrons missing from either the 1$\pi_{u_x}$ or the 1$\pi_{u_y}$ orbital.}
     \label{fig:Comp_2+_Outer_Inner_1}
\end{figure}

In Fig. \ref{fig:Comp_2+_Outer_Inner_1}, we compare the PECs we obtain for N$_2$ ion states with two electrons missing from the same orbital, inner valence or outer valence, with the respective PECs obtained in Ref. \cite{Deb}. Specifically, in Fig. \ref{fig:Comp_2+_Outer_Inner_1}, we show the PECs for the ion states with two electrons missing from the orbitals 3$\sigma_g$ (solid black), 1$\pi_{u_{x/y}}$ (solid orange), 2$\sigma_u$ (solid pink) and 2$\sigma_g$ (solid blue). The PEC of the ion state 2$\sigma_g^{-2}$ was calculated using the TS-CASSCF technique. The PECs of the 1$\pi_{u_{x/y}}^{-2}$ and 2$\sigma_u^{-2}$ ion states were calculated using the SA-CASSCF technique, since they do not have the lowest energy of their respective symmetry. However 3$\sigma_g^{-2}$ state was calculated using the CASSCF technique, since it is the lowest energy state of its respective symmetry. Fig. \ref{fig:Comp_2+_Outer_Inner_1} shows that the agreement between our results for the PECs and the ones obtained in Ref. \cite{Deb} is very good. The small disagreement for the PEC of the 2$\sigma_g^{-2}$ ion state is due to freezing different orbitals in the TS-CASSCF technique. In this work, in the first step, we freeze the orbital 2$\sigma_g$ and in the second step we freeze all the other doubly occupied orbitals. However, in Ref. \cite{Deb}, in the second step, all doubly occupied orbitals are frozen besides the 2$\sigma_g$ and 2$\sigma_u$ orbitals. The reason we freeze all occupied orbitals besides 2$\sigma_g$ in the second step is for the purpose of being consistent in our computations of the PECs, see section \ref{sec:Method} and Table \ref{tab:techniques}. \newline 

\begin{figure}[h!]
     \centering
     \includegraphics[width=\textwidth]{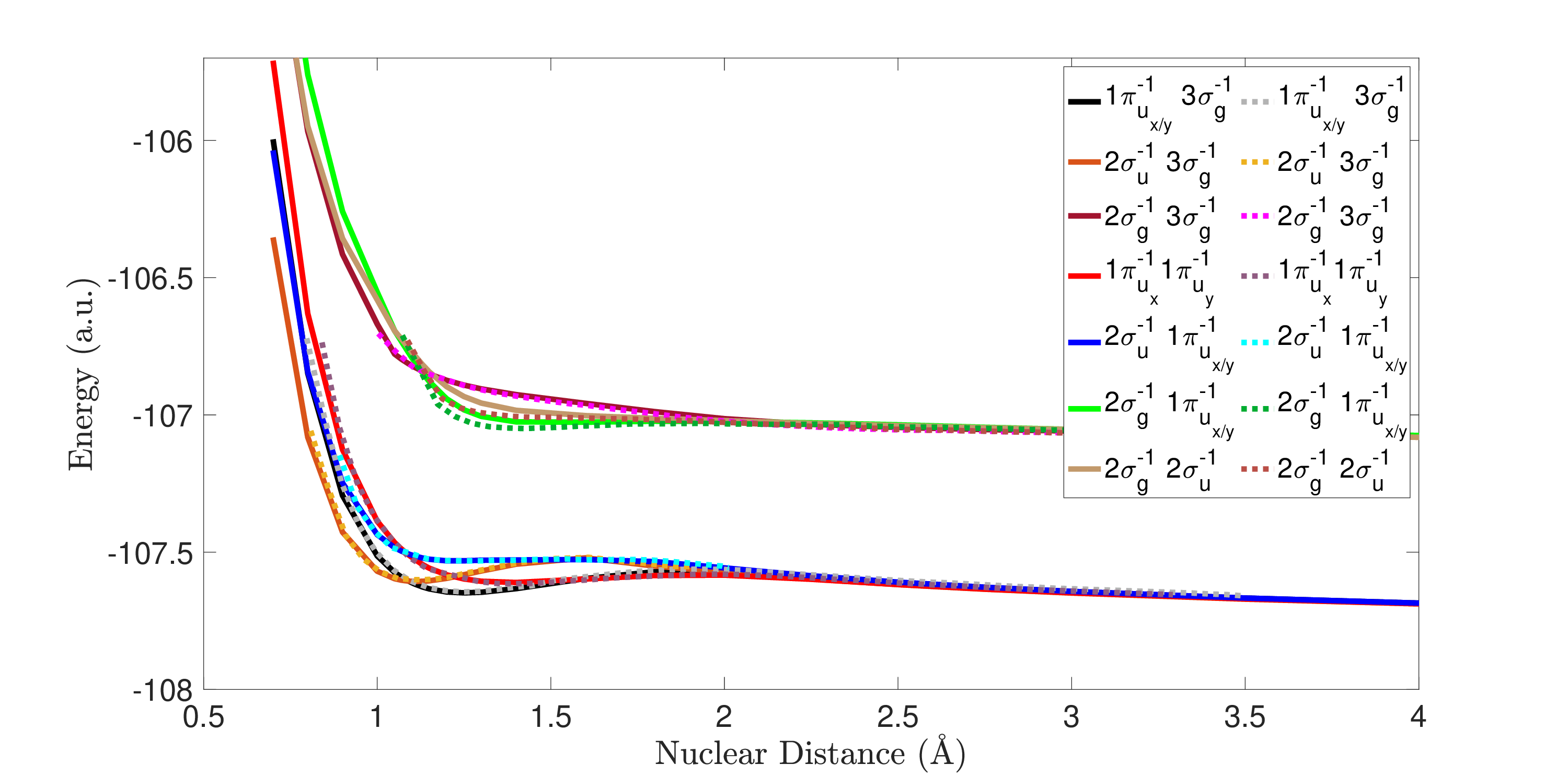}
     \caption{PECs of N$_2^{2+}$ triplet ion states, with two electrons missing from different orbitals, inner valence or outer valence ones. Solid curves represent the results obtained in this work, while dotted curves represent the results obtained in Ref. \cite{Deb}. The 1$\pi_{u_{x/y}}^{-1}$ stands for an electron missing from either the 1$\pi_{u_x}$ or the 1$\pi_{u_y}$ orbital.}
     \label{fig:Comp_2+_Outer_Inner_2}
\end{figure}

In Fig. \ref{fig:Comp_2+_Outer_Inner_2}, we compare our restuls with the ones obtained in Ref. \cite{Deb} for the doubly ionised states of N$_2$, with two electrons missing from different orbitals, inner valence or outer valence ones. In what follows, we only consider the triplet spin symmetry of the states. The PECs for the ion states 2$\sigma_g^{-1}$2$\sigma_u^{-1}$, 2$\sigma_g^{-1}$1$\pi_{u_{x/y}}^{-1}$ and 2$\sigma_g^{-1}$3$\sigma_g^{-1}$ were calculated using the TS-CASSCF technique while the remaining PECs were calculated using the CASSCF technique. We find a very good agreement between the PECs obtained in this work and the ones obtained in Ref. \cite{Deb}. A small difference is found for the PECs of the 2$\sigma_g^{-1}$2$\sigma_u^{-1}$ and 2$\sigma_g^{-1}$1$\pi_{u_{x/y}}^{-1}$ ion states. This small difference arises from freezing different orbitals in the second step. Here, we optimise the 2$\sigma_g$ orbital, while in Ref. \cite{Deb} for the 2$\sigma_g^{-1}$1$\pi_{u_{x/y}}^{-1}$ state the orbitals 2$\sigma_g$ and 1$\pi_{u_{x/y}}$ are optimised and for the 2$\sigma_g^{-1}$2$\sigma_u^{-1}$ state the orbitals 2$\sigma_g^{-1}$ and 2$\sigma_u^{-1}$ are optimised. The reason we freeze all occupied orbitals besides 2$\sigma_g$ in the second step is for the purpose of being consistent in our computations of the PECs, see section \ref{sec:Method} and Table \ref{tab:techniques}. \newline 

\begin{figure}[h!]
     \centering
     \includegraphics[width=\textwidth]{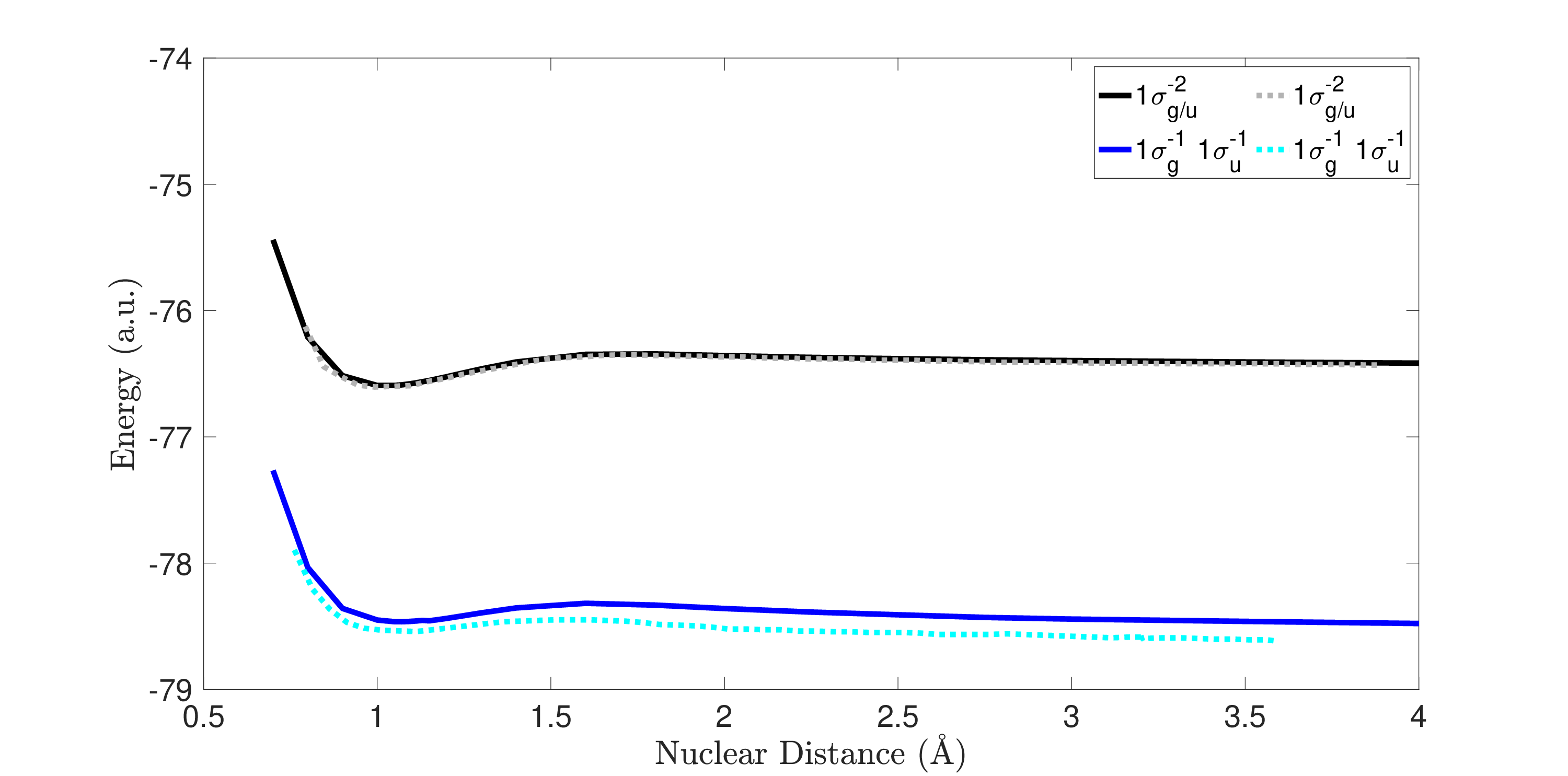}
     \caption{PECs of N$_2^{2+}$ ion states with two core electrons missing. Solid curves represent the results obtained in this work, while dotted curves represent the results obtained in Ref. \cite{Deb}. The PEC of the 1$\sigma_{g/u}^{-2}$ state, accounts for the two distinct PECs of the 1$\sigma_g^{-2}$ and 1$\sigma_u^{-2}$ states, which we find to be very similar. }
     \label{fig:Comp_2+_Core}
\end{figure}

\begin{figure}[h!]
     \centering
     \includegraphics[width=\textwidth]{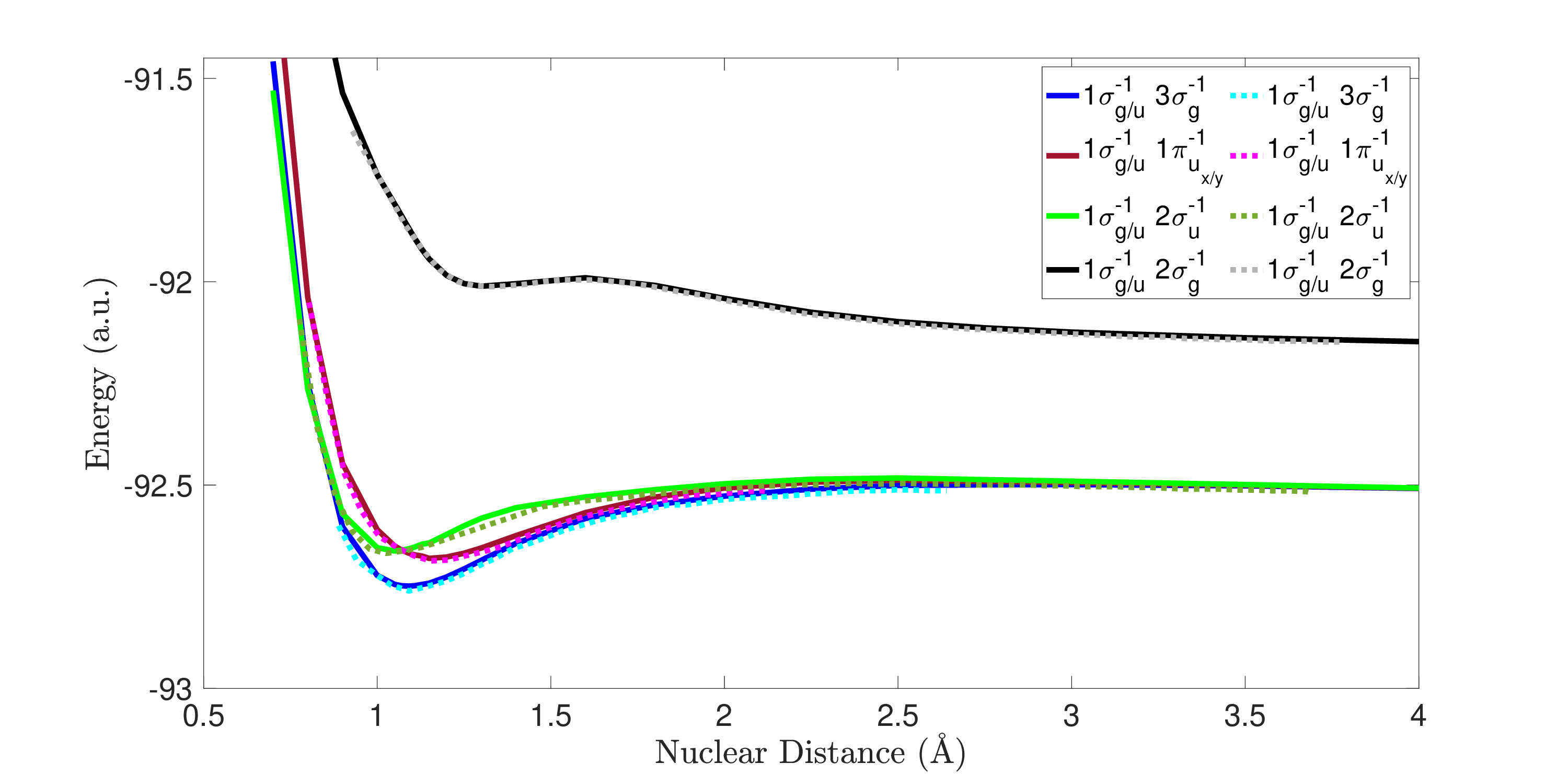}
     \caption{PECs of N$_2^{2+}$ triplet ion states with one core electron and one inner valence or outer valence electron missing. Solid curves represent the results obtained in this work, while dotted curves represent the results obtained in Ref. \cite{Deb}. The PECs including the 1$\sigma_{g/u}^{-1}$ term, account for two distinct PECs involving the 1$\sigma_g^{-1}$ and 1$\sigma_u^{-1}$, which we find to be very similar. The 1$\pi_{u_{x/y}}^{-1}$ stands for an electron missing from either the 1$\pi_{u_x}$ or the 1$\pi_{u_y}$ orbital.}
     \label{fig:Comp_2+_Core_Outer_Inner}
\end{figure}

In Fig. \ref{fig:Comp_2+_Core}, we present the PECs for N$_2$ ion states with two electrons missing from core orbitals. In Fig. \ref{fig:Comp_2+_Core_Outer_Inner} we present the PECs for the ion states with one core electron and one inner valence or outer valence electron missing. For these states, all PECs were computed using the TS-CASSCF technique, see Table \ref{tab:techniques}. Our results are found to be in very good agreement with the PECs obtained in Ref. \cite{Deb}. \newline

\subsection{PECs for triply ionised states of N$_2$}
Previous studies \cite{Bandrauk_1999,  Bocharova_2011,Wu2011} have computed the PECs for triply ionised states of N$_2$ where only outer valence electrons are missing. In what follows we present our results for the PECs of triply ionised states of N$_2$ where outer valence, inner valence and core electrons are missing. \newline

\begin{figure}[h!]
\centering
\begin{subfigure}{\textwidth}
    \includegraphics[width=\textwidth]{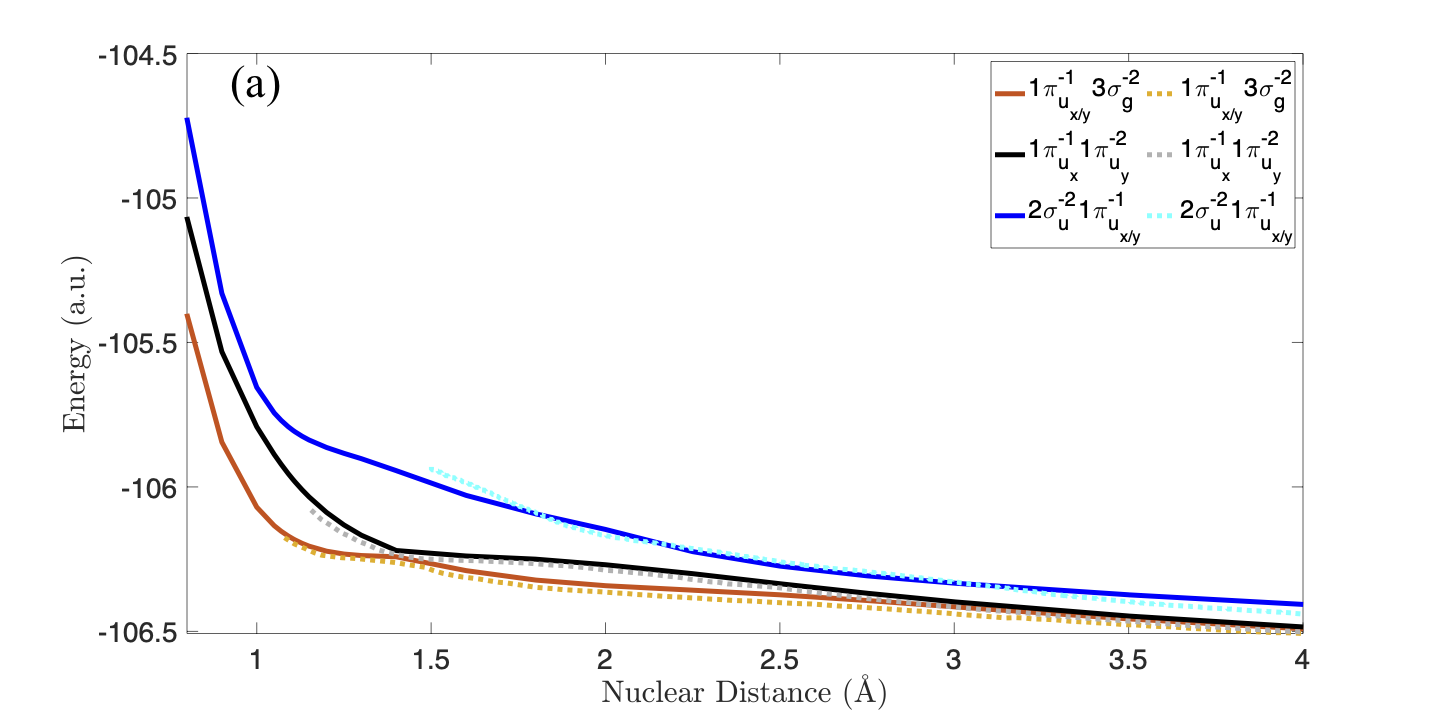}
    \label{fig:TriplyIonised_Coms_1}
\end{subfigure}
\hfill
\begin{subfigure}{\textwidth}
    \includegraphics[width=\textwidth]{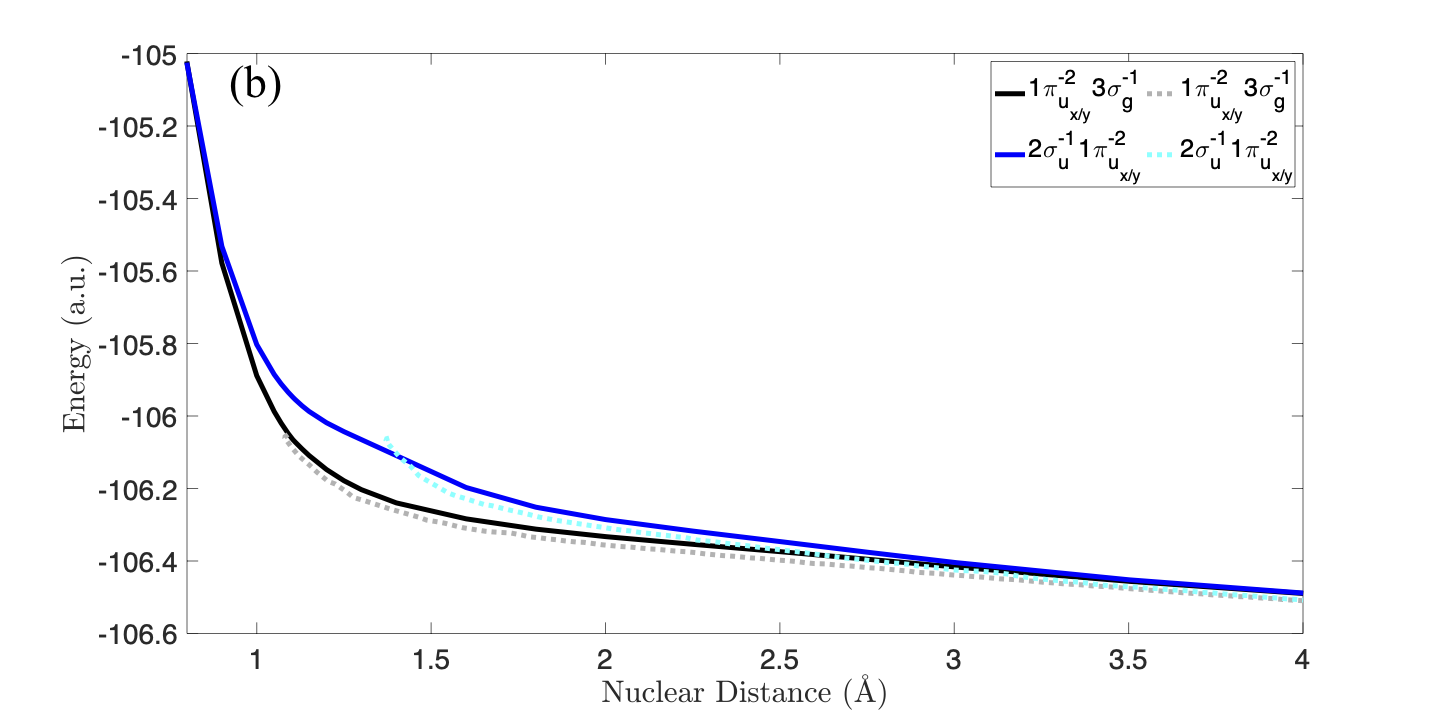}
    \label{fig:TriplyIonised_Coms_2}
\end{subfigure}
\hfill
\caption{PECs of N$_2^{3+}$ ion states with three outer valence electrons missing. Solid curves represent the results obtained in this work, while dotted curves represent the results obtained in Ref. \cite{Bandrauk_1999}. The 1$\pi_{u_{x/y}}^{-1}$ stands for an electron missing from either the 1$\pi_{u_x}$ or the 1$\pi_{u_y}$ orbital. The 1$\pi_{u_{x/y}}^{-2}$ stands for two electrons missing from either the 1$\pi_{u_x}$ or the 1$\pi_{u_y}$ orbital.}
\label{fig:TriplyIonised_Coms}
\end{figure}

First, in Fig. \ref{fig:TriplyIonised_Coms}, we compare the PECs we compute for triply ionised states of N$_2$ with outer valence electrons missing with the results obtained in Ref. \cite{Bandrauk_1999}. We find good agreement between the PECs obtained in this work and the ones obtained in Ref. \cite{Bandrauk_1999}.The small differences observed can be attributed to the different active spaces and basis sets employed. Indeed, in this work, we use an active space consisting of ten orbitals whereas in Ref. \cite{Bandrauk_1999} the active space consists of twelve orbitals.  Also, we use the  aug-cc-pVQZ basis set to compute PECs whereas in Ref. \cite{Bandrauk_1999} a different basis set is employed. \newline

\begin{figure}[h!]
     \centering
     \includegraphics[width=\textwidth]{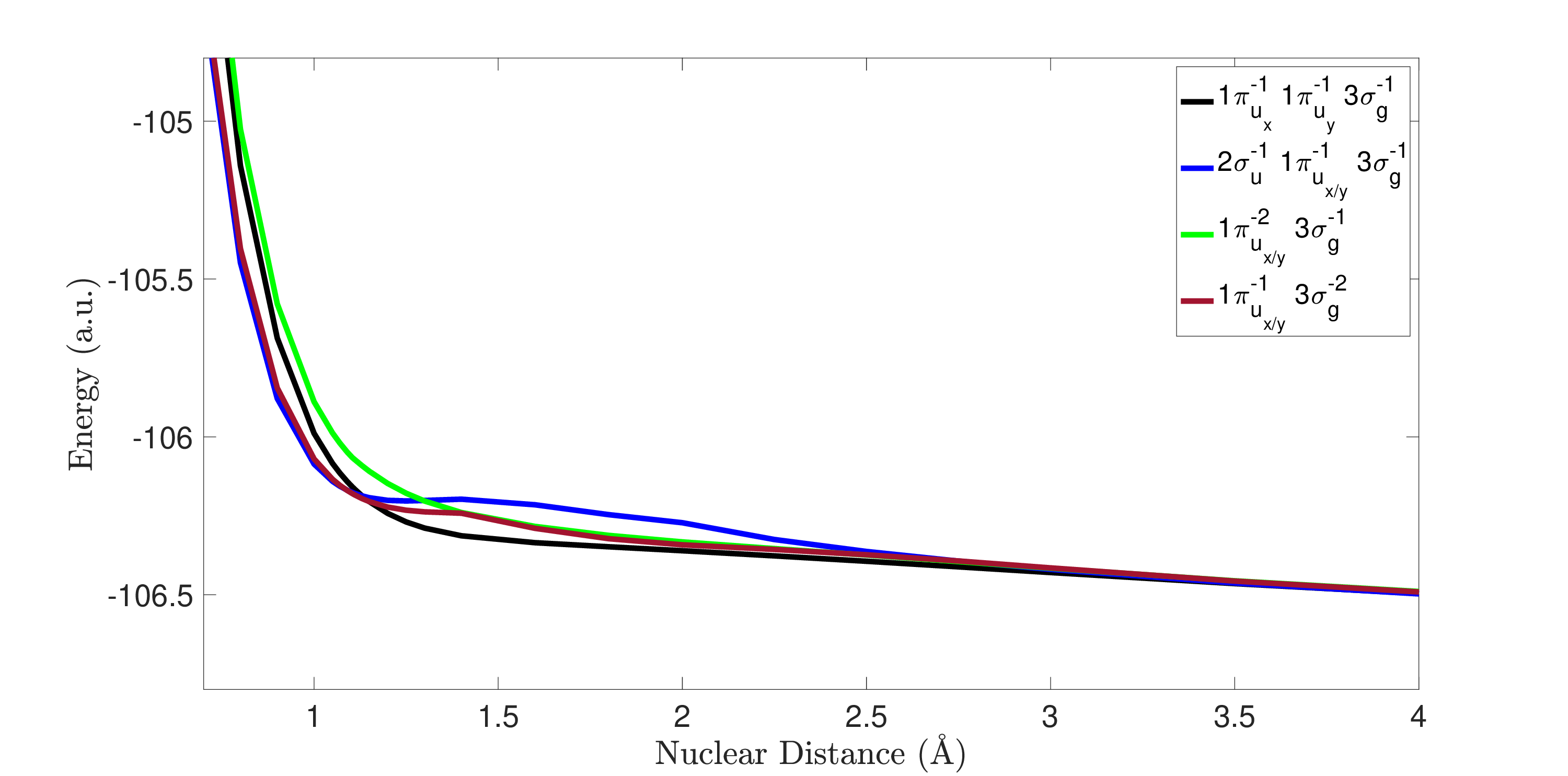}
     \caption{PECs of N$_2^{3+}$ ion states generated using the CASSCF technique. The 1$\pi_{u_{x/y}}^{-1}$ stands for an electron missing from either the 1$\pi_{u_x}$ or the 1$\pi_{u_y}$ orbital.}
     \label{fig:TriplyIonised_CASSCF}
\end{figure}

\begin{figure}[h!]
     \centering
     \includegraphics[width=\textwidth]{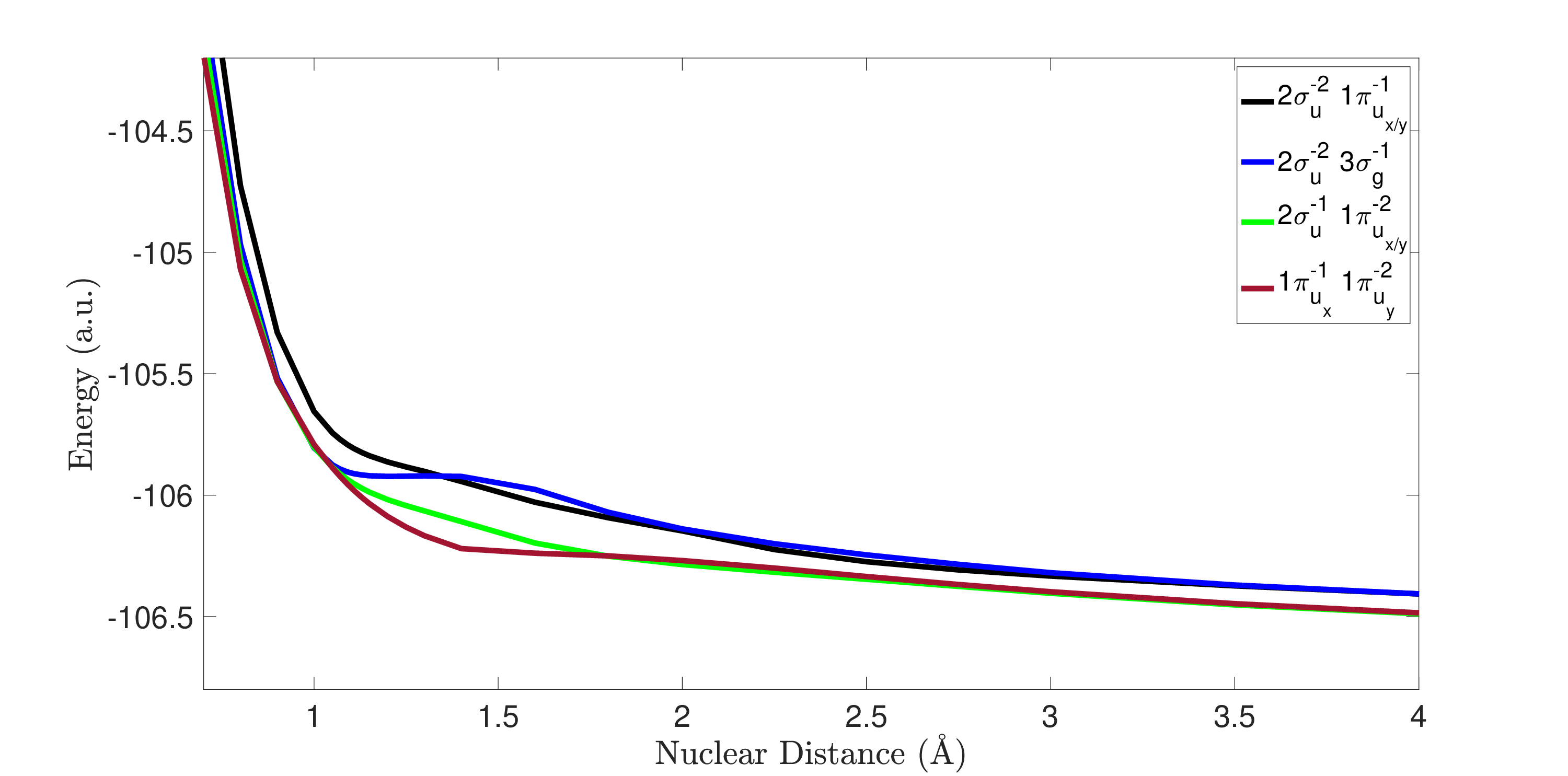}
     \caption{PECs of N$_2^{3+}$ ion states generated using the SA-CASSCF technique. The 1$\pi_{u_{x/y}}^{-1}$ stands for an electron missing from either the 1$\pi_{u_x}$ or the 1$\pi_{u_y}$ orbital. The 1$\pi_{u_{x/y}}^{-2}$ stands for two electrons missing from either the 1$\pi_{u_x}$ or the 1$\pi_{u_y}$ orbital.}
     \label{fig:TriplyIonised_SA-CASSCF}
\end{figure}

\begin{figure}[h!]
     \centering
     \includegraphics[width=\textwidth]{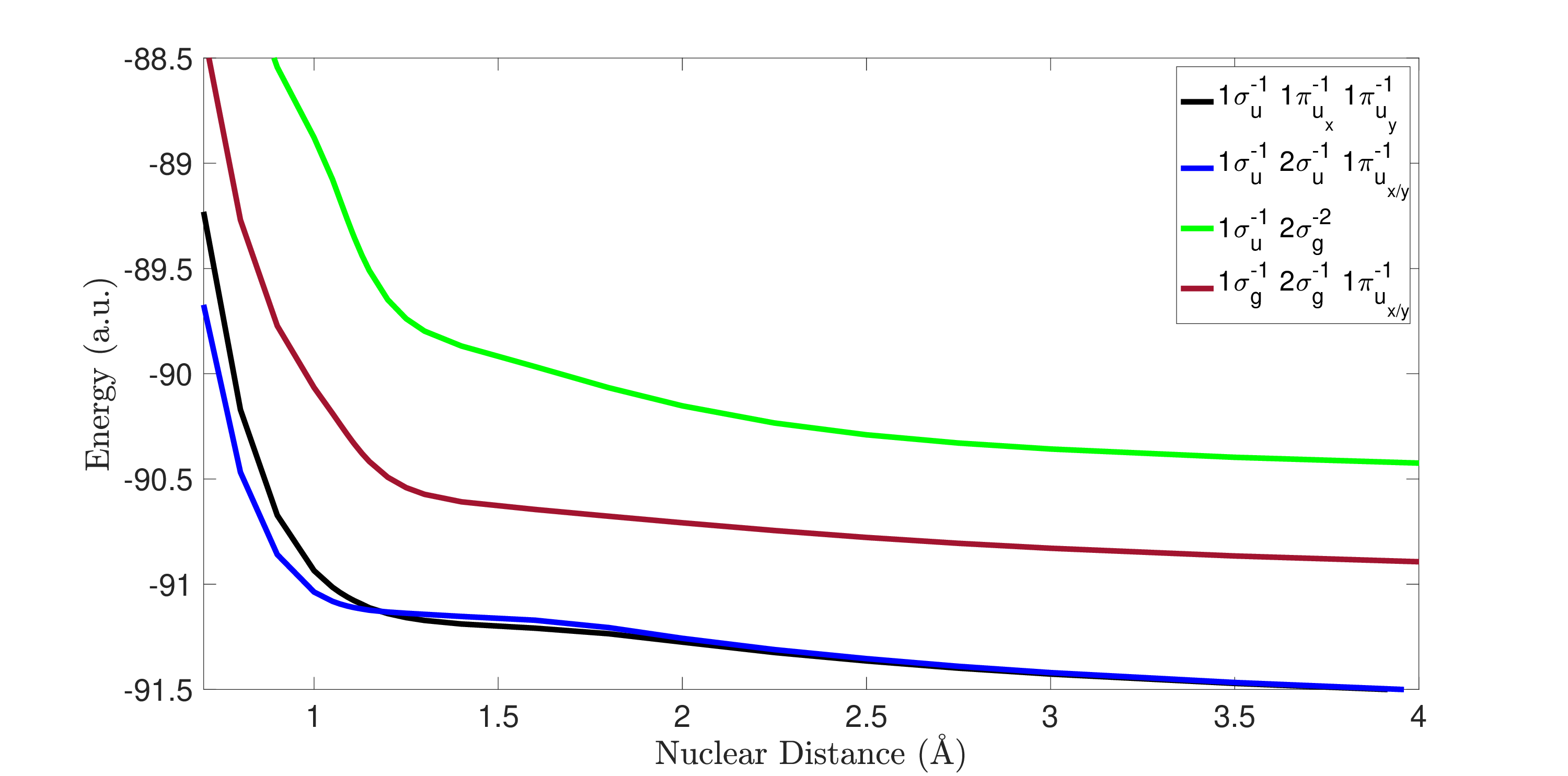}
     \caption{PECs of N$_2^{3+}$ ion states generated using the TS-CASSCF technique. The 1$\pi_{u_{x/y}}^{-1}$ stands for an electron missing from either the 1$\pi_{u_x}$ or the 1$\pi_{u_y}$ orbital.}
     \label{fig:TriplyIonised_TS-CASSCF}
\end{figure}

\begin{figure}[h!]
     \centering
     \includegraphics[width=\textwidth]{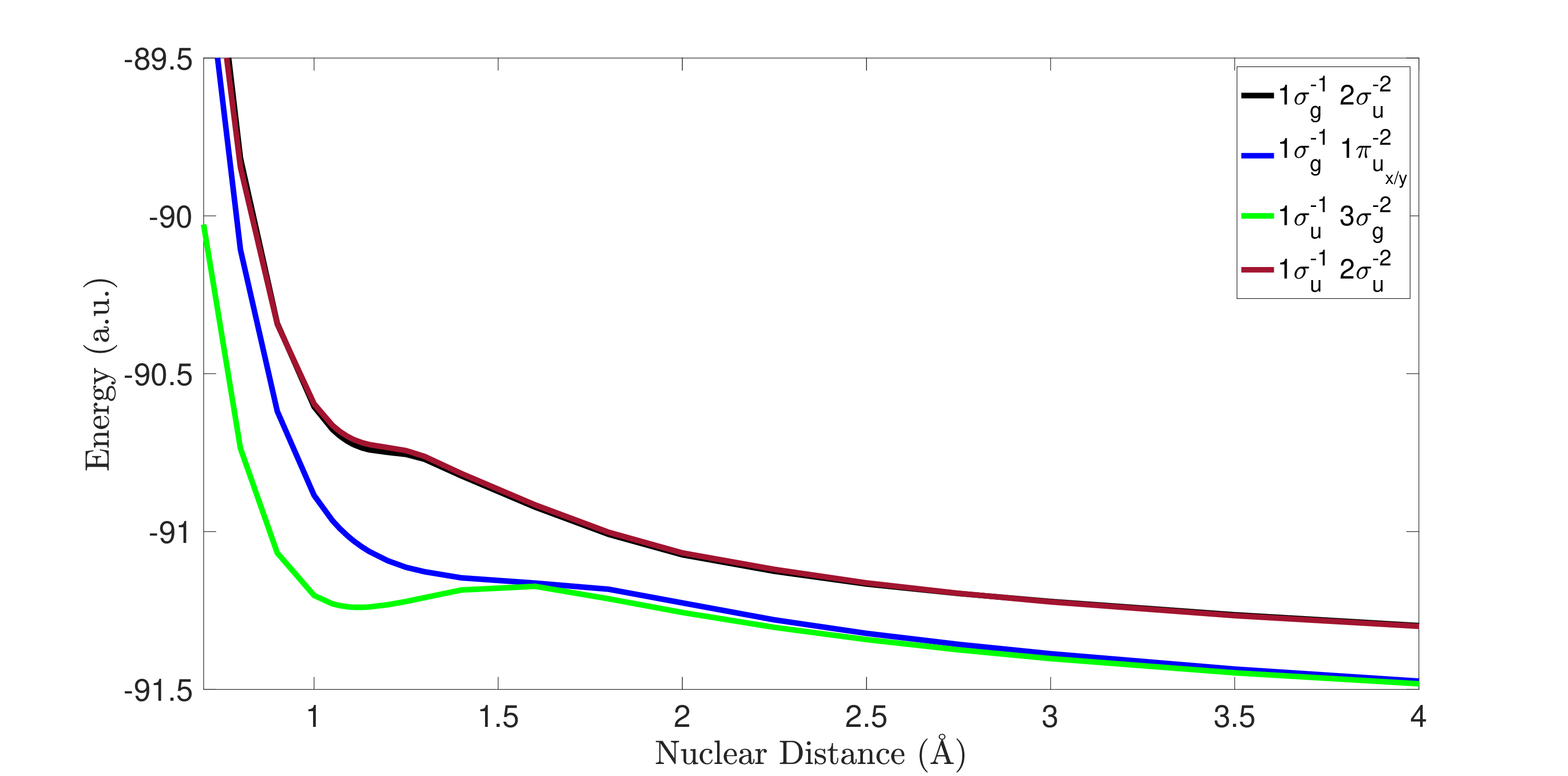}
     \caption{PECs of N$_2^{3+}$ ion states generated using the SA-TS-CASSCF technique. The 1$\pi_{u_{x/y}}^{-2}$ stands for two electrons missing from either the 1$\pi_{u_x}$ or the 1$\pi_{u_y}$ orbital.}
     \label{fig:TriplyIonised_SA-TS-CASSCF}
\end{figure}

Next, we obtain the PECs for the triply ionised states of N$_2$ for different combinations of outer valence, inner valence and core electrons missing. We group the PECs from Fig. \ref{fig:TriplyIonised_CASSCF} to Fig. \ref{fig:TriplyIonised_SA-TS-CASSCF} according to the technique we use to compute these PECs. The CASSCF and SA-CASSCF techniques are used to calculate states with 3 electrons missing from outer valence orbitals. For these latter ion states, we employ the SA-CASSCF technique when the ion state is not the lowest energy state of its respective symmetry. For triply ionised N$_2$ states with an electron missing from a core or inner valence orbital, we employ the TS-CASSCF and SA-TS-CASSCF techniques. The SA-TS-CASSCF technique is employed when the state is not the lowest energy state of its respective symmetry. \newline

In Fig.  \ref{fig:TriplyIonised_CASSCF}, we present the PECs of triply ionised states computed using the CASSCF technique. We find the PECs of ion states with outer valence electrons missing to be repulsive, leading to molecular dissociation. This is also shown in Fig. \ref{fig:TriplyIonised_Coms} and Ref. \cite{Bandrauk_1999} where all PECs computed for triply ionised states of N$_2$ with outer valence electrons missing are found to be repulsive and lead to Coulomb explosion. The four PECs shown in Fig. \ref{fig:TriplyIonised_CASSCF} have the same dissociation limit. In Fig.  \ref{fig:TriplyIonised_SA-CASSCF}, we show the PECs of triply ionised states with outer valence electrons missing computed using the SA-CASSCF technique. The PECs shown are found to be repulsive with the exception of state 2$\sigma_u^{-2}$3$\sigma_g^{-1}$ which has a minimum close to 1.20 $\AA$.  \newline

In Fig. \ref{fig:TriplyIonised_TS-CASSCF}, we obtain the PECs of triply ionised states computed using the TS-CASSCF technique. All the PECs shown in Fig. \ref{fig:TriplyIonised_TS-CASSCF} have one core and two inner or outer valence electrons missing and are repulsive leading to dissociation.  Finally, in Fig. \ref{fig:TriplyIonised_SA-TS-CASSCF}, we obtain the PECs of triply ionised states computed using the SA-TS-CASSCF technique. The states shown are repulsive leading to molecular dissociation, with the exception of state 1$\sigma_u^{-1}$3$\sigma_g^{-2}$ which has a minimum at 1.13 $\AA$.  \newline

Moreover, the states 1$\sigma_g^{-1}$2$\sigma_u^{-2}$ and 1$\sigma_u^{-1}$2$\sigma_u^{-2}$ are similar in shape with the 1$\sigma_u^{-1}$2$\sigma_u^{-2}$ being slightly higher in energy. We have identified a similar pattern for doubly ionised states of N$_2$ when one of the two electrons missing is from the 1$\sigma_u$ or 1$\sigma_g$ orbital, see Ref. \cite{Deb}.

\newpage

\subsection{PECs for quadruply ionised states of N$_2$}

\begin{figure}[h!]
\centering
\includegraphics[width=\textwidth]{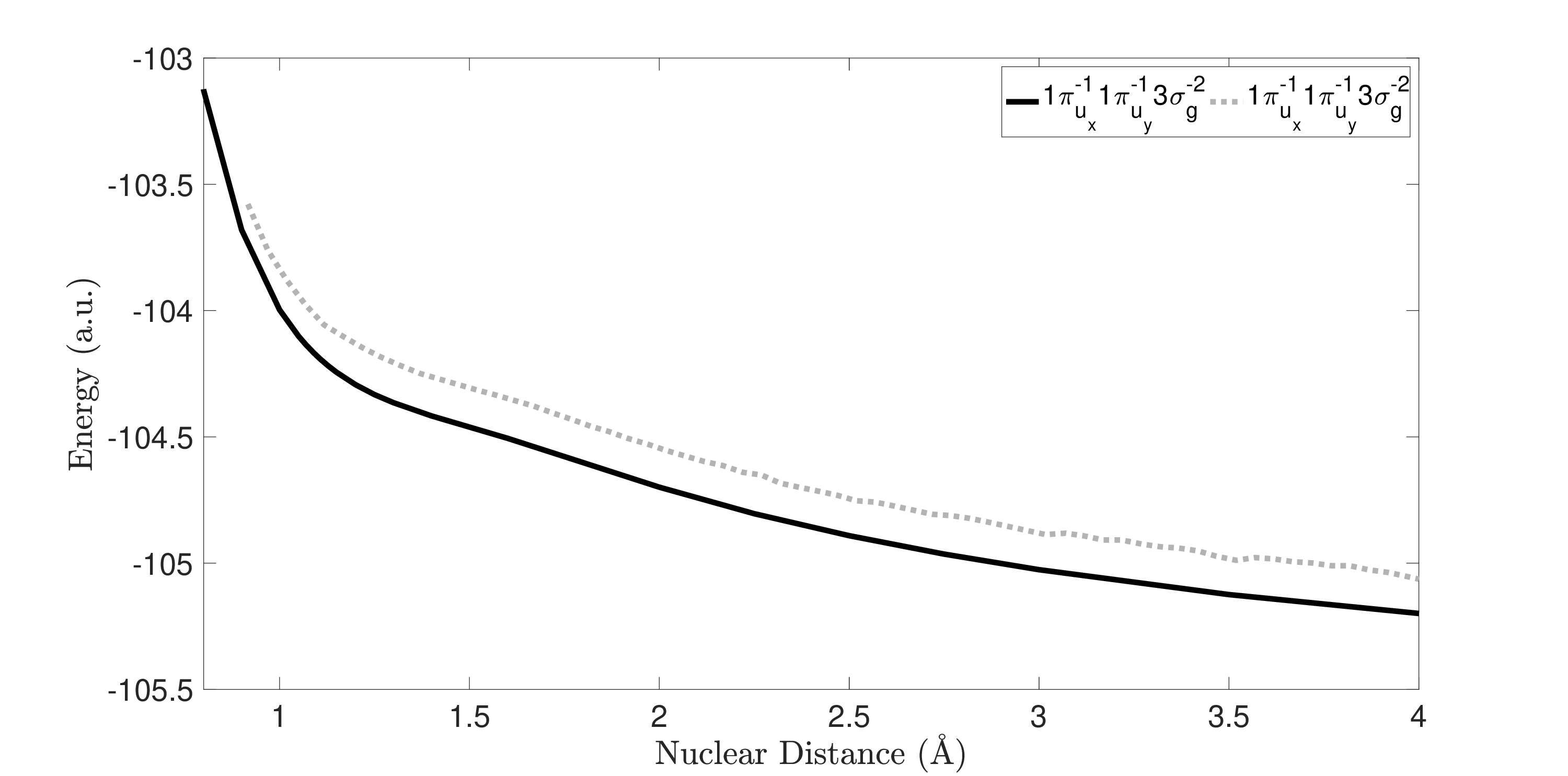} 
\caption{PEC of the triplet 1$\pi_{u_{x}}^{-1}$1$\pi_{u_{y}}^{-1}$3$\sigma_g^{-2}$ N$_2^{4+}$ ion state with four outer valence electrons missing. Solid curve represents the result obtained in this work, while the dotted curve represents the result obtained in Ref. \cite{Wu2011}.}
\label{fig:QuadIonised_Coms}
\end{figure}

A previous study \cite{Wu2011} has obtained the PEC of the triplet 1$\pi_{u_{x}}^{-1}$1$\pi_{u_{y}}^{-1}$3$\sigma_g^{-2}$ N$_2^{4+}$ state, where the electrons missing are from outer valence orbitals. This state was obtained using the multiconfiguration second-order perturbation theory (CASPT2) method \cite{CASPT2_1,CASPT2_2}. In Fig. \ref{fig:QuadIonised_Coms}, we obtain the PEC for the triplet 1$\pi_{u_{x}}^{-1}$1$\pi_{u_{y}}^{-1}$3$\sigma_g^{-2}$ state and compare with the result in Ref. \cite{Wu2011}. We find  that our PEC agrees well with the one obtained in Ref. \cite{Wu2011}, while the differences observed are due to the different methods and basis sets employed to generate the PEC of the N$_2^{4+}$ state.  \newline

\begin{figure}[h!]
     \centering
     \includegraphics[width=\textwidth]{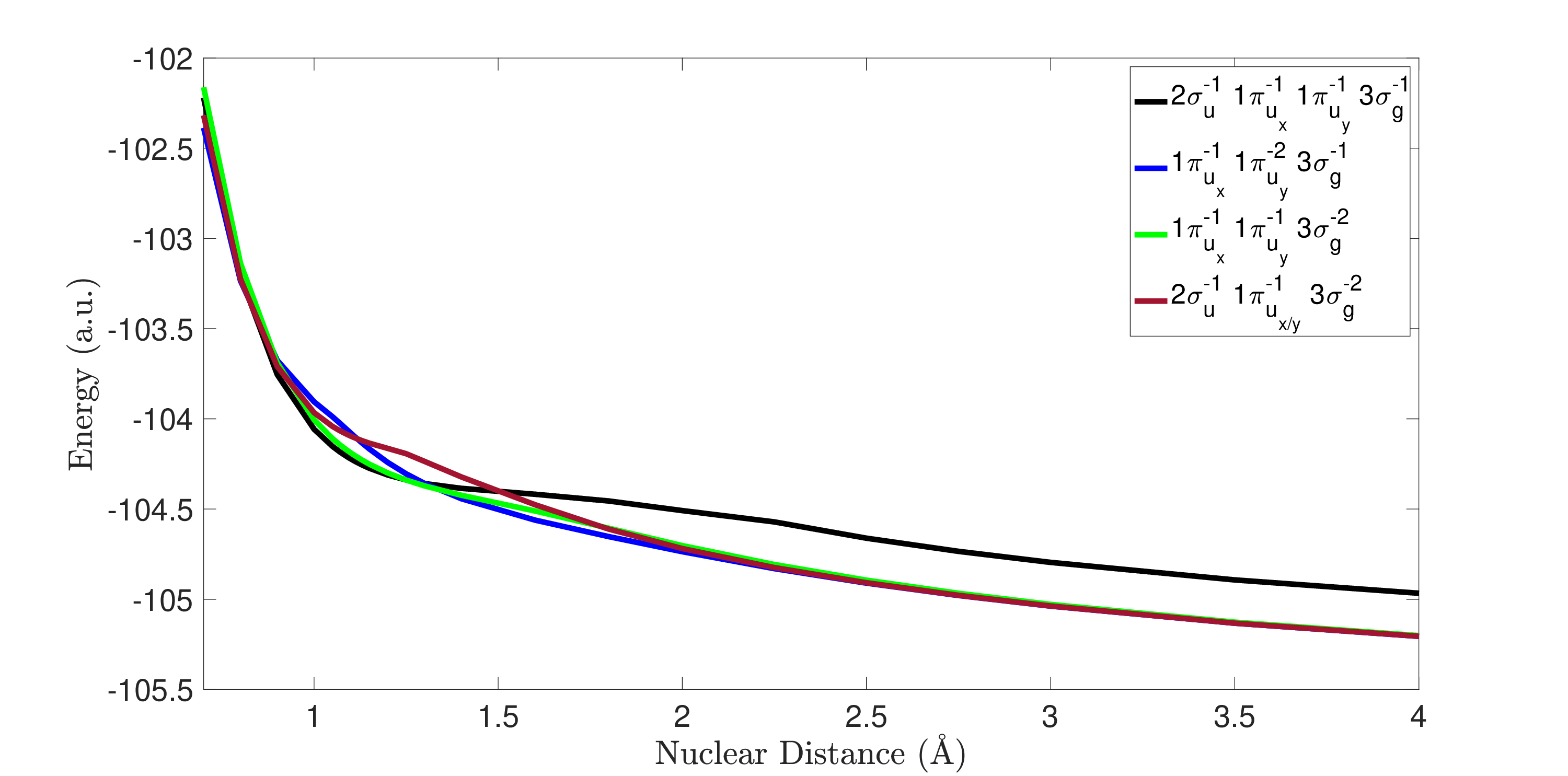}
     \caption{PECs of N$_2^{4+}$ ion states generated using the CASSCF technique.  The 1$\pi_{u_{x/y}}^{-1}$ stands for an electron missing from either the 1$\pi_{u_x}$ or the 1$\pi_{u_y}$ orbital.}
     \label{fig:QuadIonised_CASSCF}
\end{figure}

\begin{figure}[h!]
     \centering
     \includegraphics[width=\textwidth]{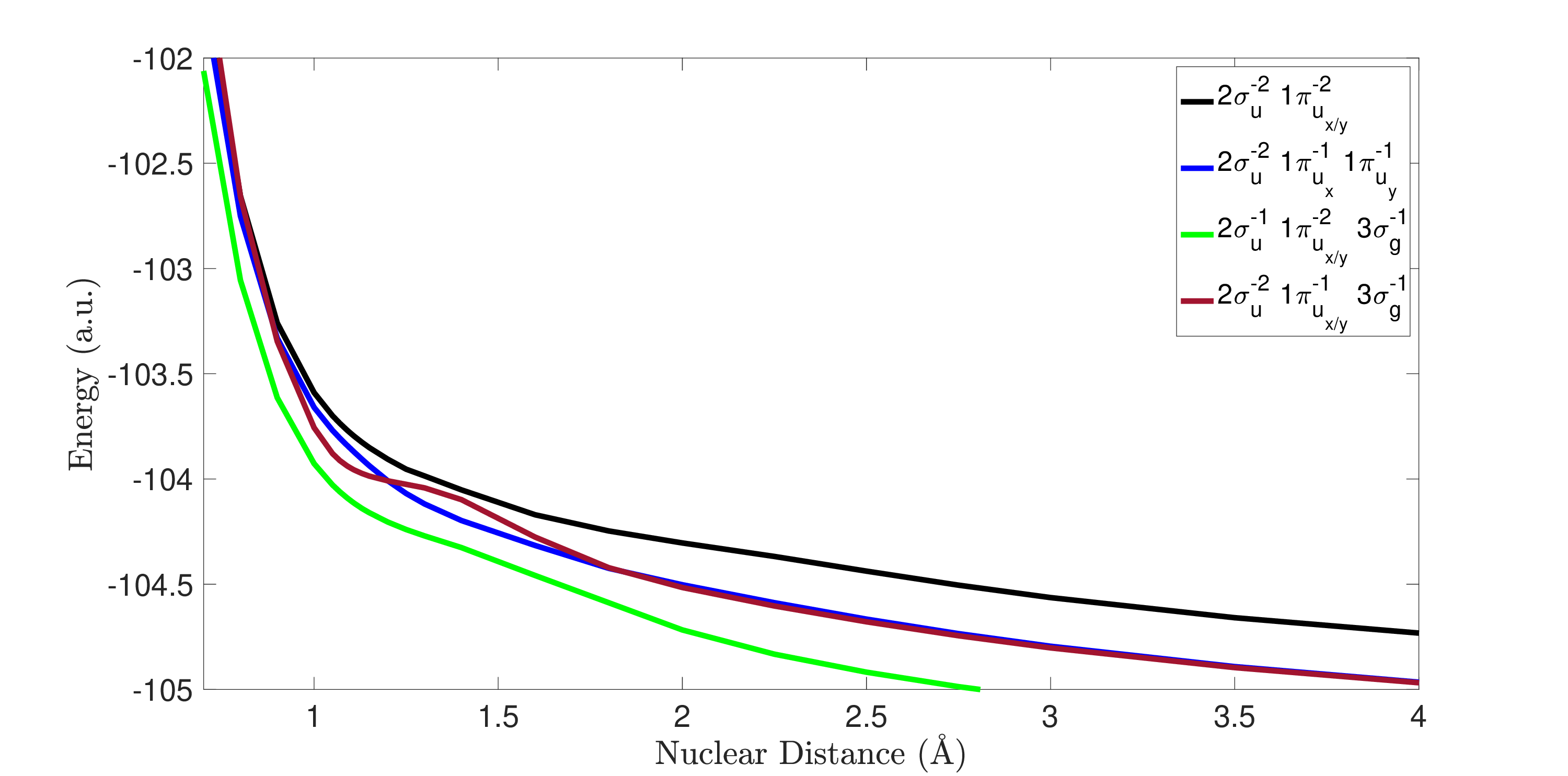}
     \caption{PECs of N$_2^{4+}$ ion states generated using the SA-CASSCF technique. The 1$\pi_{u_{x/y}}^{-1}$ stands for an electron missing from either the 1$\pi_{u_x}$ or the 1$\pi_{u_y}$ orbital. The 1$\pi_{u_{x/y}}^{-2}$ stands for two electrons missing from either the 1$\pi_{u_x}$ or the 1$\pi_{u_y}$ orbital.}
     \label{fig:QuadIonised_SA-CASSCF}
\end{figure}

\begin{figure}[h!]
     \centering
     \includegraphics[width=\textwidth]{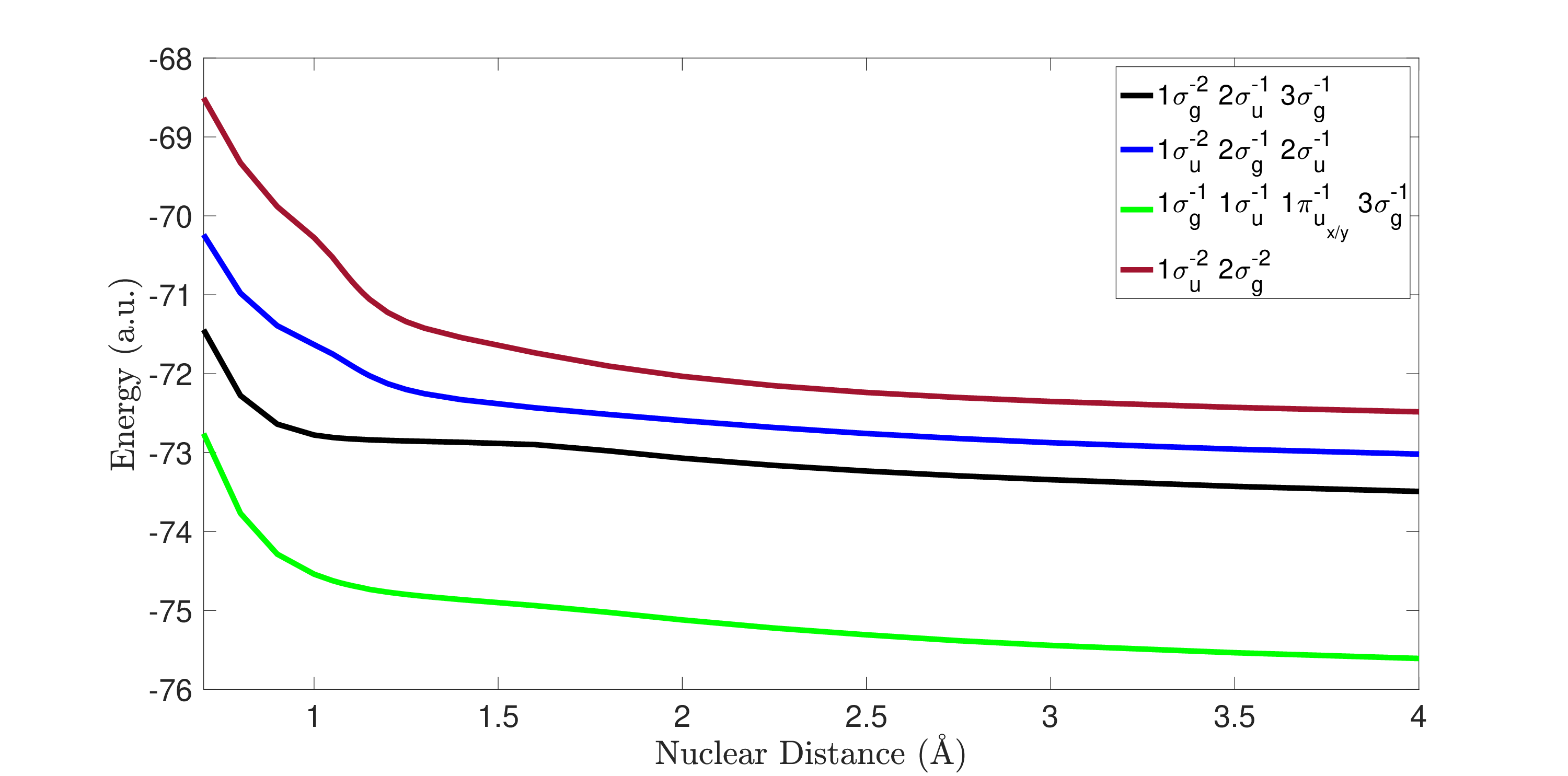}
     \caption{PECs of N$_2^{4+}$ ion states generated using the TS-CASSCF technique. The 1$\pi_{u_{x/y}}^{-1}$ stands for an electron missing from either the 1$\pi_{u_x}$ or the 1$\pi_{u_y}$ orbital. }
     \label{fig:QuadIonised_TS-CASSCF}
\end{figure}

\begin{figure}[h!]
     \centering
     \includegraphics[width=\textwidth]{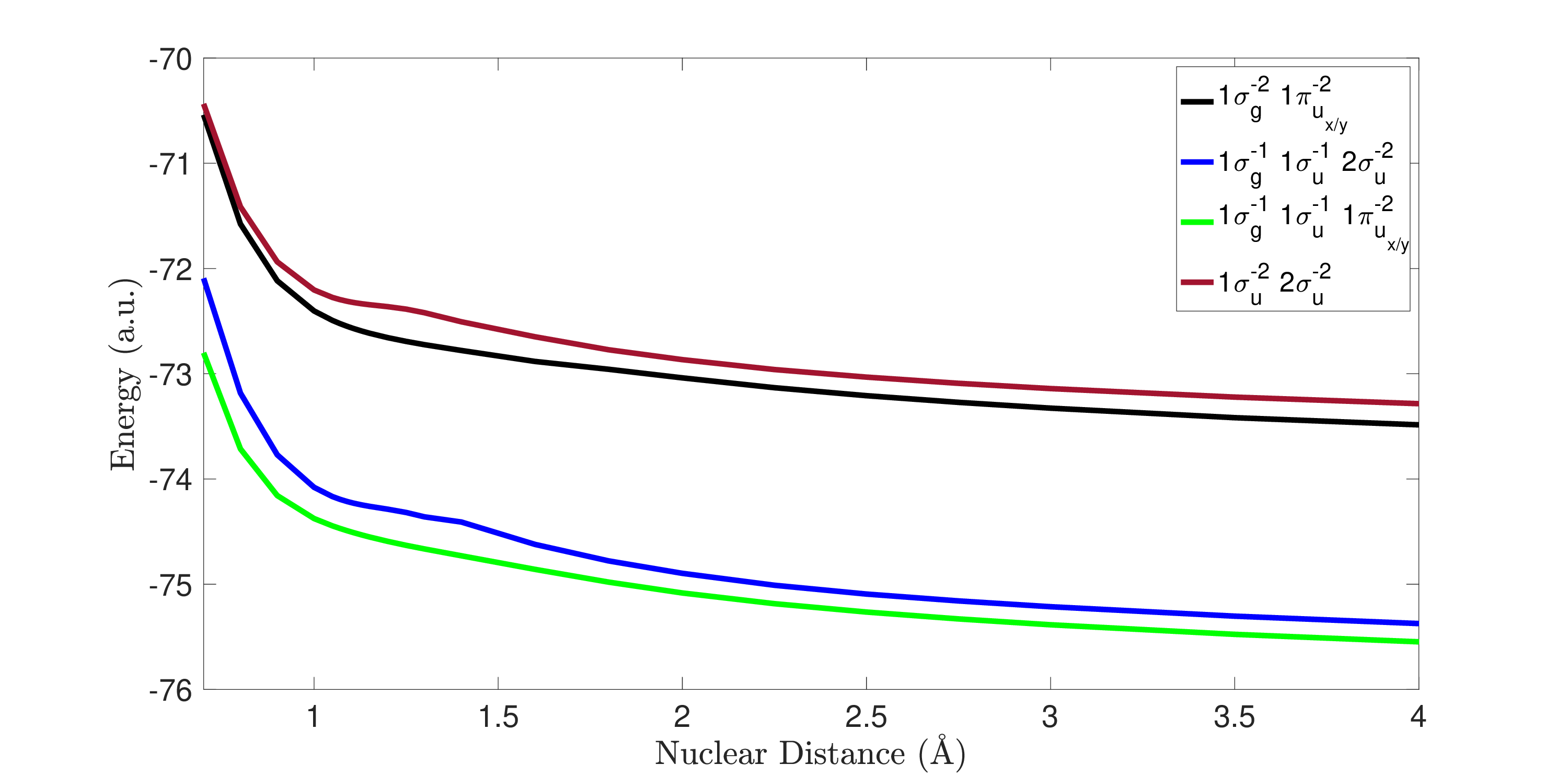}
     \caption{PECs of N$_2^{4+}$ ion states generated using the SA-TS-CASSCF technique. The 1$\pi_{u_{x/y}}^{-2}$ stands for two electrons missing from either the 1$\pi_{u_x}$ or the 1$\pi_{u_y}$ orbital. }
     \label{fig:QuadIonised_SA-TS-CASSCF}
\end{figure}

In what follows, we obtain the PECs of quadruply ionised states of N$_2$ where outer valence, inner valence and core electrons are missing. We group the PECs from Fig. \ref{fig:QuadIonised_CASSCF} to Fig. \ref{fig:QuadIonised_SA-TS-CASSCF} according to the technique we use to compute them. The CASSCF and SA-CASSCF techniques are used to calculate states with four electrons missing from outer valence orbitals. For these latter ion states, we employ the SA-CASSCF technique when the ion state is not the lowest energy state of its respective symmetry. For quadruply ionised N$_2$ states with an electron missing from a core or inner valence orbital, we employ the TS-CASSCF and SA-TS-CASSCF techniques. The SA-TS-CASSCF technique is employed when the state is not the lowest energy state of its respective symmetry. \newline

In Fig. \ref{fig:QuadIonised_CASSCF}, we present the PECs of quadruply ionised states with outer valence electrons missing computed using the CASSCF technique. We find these PECs to be repulsive, leading to molecular dissociation. We also note that these PECs for quadruply ionised states of N$_2$ decrease more rapidly as a function of distance compared to the PECs of triply ionised states with three outer valence electrons missing shown in Fig. \ref{fig:TriplyIonised_CASSCF}. This is consistent with the greater Coulomb repulsion taking place between the two cores during dissociation of the N$_2^{4+}$ ions.  \newline 

In Fig. \ref{fig:QuadIonised_SA-CASSCF}, we show the PECs of quadruply ionised states of N$_2$ with outer valence electrons missing computed using the SA-CASSCF technique. The PECs shown are found to be repulsive leading to molecular dissociation. Similar to the PECs in Fig. \ref{fig:QuadIonised_CASSCF}, these quadruply ionised states decrease more rapidly as a function of distance when compared to PECs with only three outer valence electrons missing shown in Fig. \ref{fig:TriplyIonised_SA-CASSCF}.  \newline

In Fig.\ref{fig:QuadIonised_TS-CASSCF}, we obtain the PECs of quadruply ionised states, with two core holes. These PECs are computed using the TS-CASSCF technique and are repulsive leading to molecular dissociation. In Fig. \ref{fig:QuadIonised_SA-TS-CASSCF}, we obtain the PECs of quadruply ionised states, with two core holes. These PECs are computed using the SA-TS-CASSCF technique and are repulsive leading to molecular dissociation. Finally, we note that our finding of N$_2^{4+}$ ion states being repulsive and leading to molecular dissociation is in agreement with previous studies \cite{Wu2011, Codling1991, Baldit2005, Safvan1994}. 

\section{Conclusions}
We have obtained the potential energy curves for molecular ions up to N$_2^{4+}$. We have identified the specific techniques employed to calculate the PECs of N$_2$ ion states depending on how many electrons are missing and the specific combination of outer valence, inner valence and core electrons missing.All these techniques are based on the multi configurational self-consistent field method. Also, we provide the python code used to generate the potential energy curves of all ions of N$_2$ with up to four electrons missing. The code can be modified to calculate the PECs of higher charged ion states of N$_2$, as well as the PECs of different molecules using the same techniques discussed in this study. 

\section*{References}

\bibliography{mybibfile}

\end{document}